\documentclass[11pt]{elsarticle}

\pdfoutput=1

\makeatletter
\def\ps@pprintTitle{%
  \let\@oddhead\@empty
  \let\@evenhead\@empty
  \let\@oddfoot\@empty
  \let\@evenfoot\@oddfoot
}
\makeatother
\usepackage{amsmath}
\usepackage{url}
\usepackage{breakurl}
\usepackage[breaklinks,
            colorlinks = true,
            linkcolor = blue,
            urlcolor  = blue,
            citecolor = blue,
            anchorcolor = blue]{hyperref}

\usepackage{lineno}
\usepackage{comment}
\usepackage{graphicx}
\usepackage[margin=1.25in]{geometry}
\usepackage[usenames,dvipsnames]{color}

\newcommand\acp{\begin{center}
\rule[-0.2in]{\hsize}{0.01in}\\\rule{\hsize}{0.01in}\\
\vskip 0.1in Submitted to the  Proceedings\\ 
of the African Conference on Fundamental and Applied Physics
    \vskip 0.05in
    {\it Second Edition, ACP2021, March 7--11, 2022 --- Virtual Event}\\
\rule{\hsize}{0.01in}\\\rule[+0.2in]{\hsize}{0.01in} \\
\end{center}}

\usepackage[firstpage=true]{background}
\backgroundsetup{contents={\parbox{6.5in}{\acp}}, scale=1,placement=top,opacity=1,color=black,position={3.25in,1.2in}}

\usepackage{fancyhdr}
\fancypagestyle{plain}{%
  \fancyhf{}%
  \fancyhead[C]{}
  \fancyfoot[C]{\thepage}
}

\fancypagestyle{empty}{%
  \fancyhf{}%
  \fancyhead[C]{{\it ACP2021, March 7--11, 2022 --- Virtual Edition}}
  \fancyfoot[C]{\thepage}
}
\begin{document}

\begin{frontmatter}


\title{Gluon mass from charmonium radiative decay and its impact on the hybrid meson properties}

\author[add1]{Azzeddine Benhamida\corref{cor1}}
\ead{benhmidaazou@gmail.com}
\author[add2]{Lahouari Semlala}

\cortext[cor1]{Corresponding Author}

\address[add1]{Faculty of Exact Science, LPTO, University of Oran 1-Ahmed Ben Bella, Algeria}
\address[add2]{Ecole Supérieure en Génie Electrique et Energétique d’Oran, Algeria}

\begin{abstract}
\noindent 
We study the photon spectrum in the radiative decays of the charmonium $\frac{1}{\Gamma}\frac{d\Gamma (J/\psi\rightarrow\gamma gg)}{dz}$ as an excellent scene for hunting the exotic mesons (hybrids $q\bar{q}g$ and glueballs $gg$) and an ideal frame for the theoretical and phenomenological investigations of the gluon nature in IR, by subtle estimation of its non-zero effective mass which seems required in Quark-Gluon Constituent Model (QGCM) for identifying the nature of these exotic species.
\end{abstract}


\end{frontmatter}

%



\section{Introduction}
\label{sec:intro}
\begin{figure}[!htbp]
\begin{center}
\includegraphics[width=\textwidth]{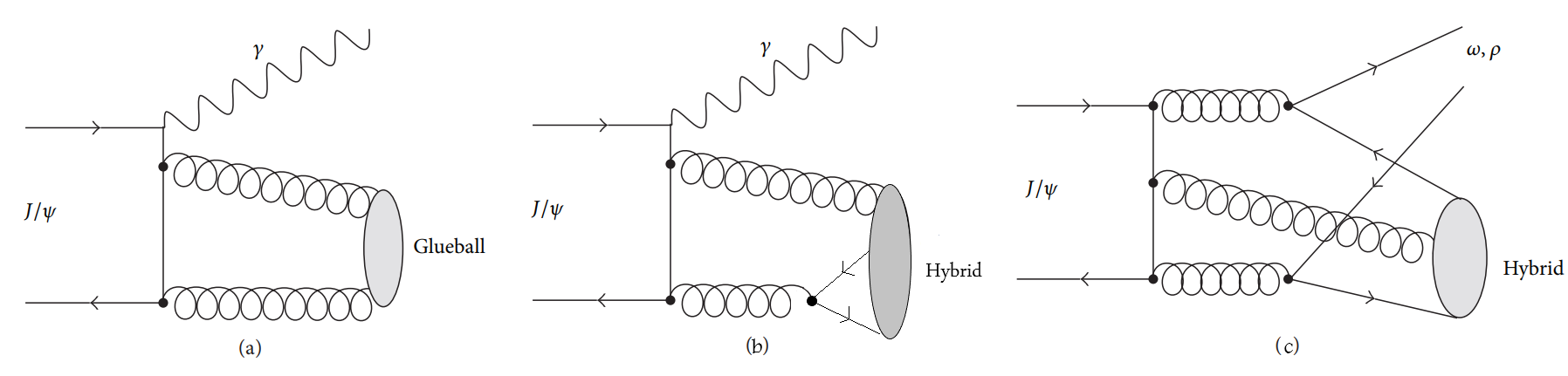}
\end{center}
\caption{(a, b) Glueball and hybrid meson production in the decay $J/\psi\rightarrow \gamma (X_G,X_H)$, (c) a production of hybrid meson through $J/\psi\rightarrow (\omega, \rho) X_H$.}
\label{fig-quarkonium}
\end{figure}

\noindent
The radiative decay of heavy quarkonium $J/\psi\rightarrow \gamma gg$ is the most promising process to search for glueballs $gg$ and to a lesser extent hybrids $q\bar{q}g$. Its advantage is that the $c\bar{c}$ system decay into light quarks is suppressed by the OZI rule \cite{wu_charmonium_2019}
. Therefore, in most decays, the $J/\psi$ undergoes a transition into three gluons which then decay to form hadrons. But $J/\psi$ can also decay into two gluons and a photon. The photon can be experimentally detected, while the two gluons ``theoretically'' can bind together to form a glueball (see Figure \ref{fig-quarkonium}), or form a hybrid via quark pair creation (QPC) by the annihilation of one of the gluons, and decay subsequently to light exotic mesons.

The main salient feature of these exotic mesons is the excitation gluon picture, as it is presumed by many potential models as a massive quasi-constituent particle ( see \cite{benhamida_hybrid_2020} and references therein) as it plays a double role inside the hadron; since it propagates the interaction between color sources (valence quarks) and, being itself colored it undergoes the interaction.  But the exploration remains neither complete nor definitive since there is a slight problem with the concept “constituent”. Furthermore, it is not clear yet whether the number of gluons is conserved, and do they have a non-vanishing rest mass in the infrared (IR) \cite{cornwall_dynamical_1982}.

\section{Phenomenological gluon mass in $J/\psi$ radiative decay}
\label{sec:grate}
\noindent
First we simulate the photon spectrum in $J/\psi\rightarrow \gamma gg$ to extract $\alpha_s$ from the ratio
$$R_{\gamma}=\frac{\Gamma_{\gamma gg}}{\Gamma_{ggg}}=\frac{N_{\gamma gg}}{N_{ggg}} \quad \sim \frac{\alpha}{\alpha_s}$$
(here $N_{\gamma gg}$ and $N_{ggg}$ represent the number of event) and then estimate the Phenomenological gluon mass by fitting the photon spectrum $\frac{1}{\Gamma}\frac{d\Gamma}{dz}$ (where $z=2E_{\gamma}/M_{J/\psi}$) taking into account all possible theoretical corrections:

\subsection*{relativistic corrections}
Since heavy quarkonium contains at least heavy (c or b) quark, the simulations using non-relativistic QCD or heavy quark effective theory, which expanses the QCD Lagrangian in powers of the heavy quark velocity, or the heavy quark mass.

The NRQCD factorization involves separating the short-distance physics of the annihilation process at scales $\sim 1/m$ that are calculated perturbatively in $\alpha_s(m_q)$ and encoded in matching coefficients, from the non-relativistic physics of the quarkonium system over scales $\sim 1/(mv)$ ($v$ being the heavy-quark velocity in the quarkonium rest frame).

The long-distance effects are parameterized by a nonperturbative matrix element $\langle \mathcal{O}^V_n\rangle$ which is subject to the power counting rules of NRQCD and can be evaluated numerically using (NRQCD) lattice simulations.

\subsection*{High order QCD corrections}
The  perturbative HO QCD correction basically split into two types of contributions, direct and fragmentation \cite{tormo_inclusive_2007} 
;
$$\frac{d\Gamma}{dz}=\frac{d\Gamma^{dir}}{dz}+\frac{d\Gamma^{frag}}{dz}$$

where

$$\frac{d\Gamma^{dir}}{dz}=\sum_i C_i(M,z)\langle \Psi_{J/\psi}|\mathcal{O}_i|\Psi_{J/\psi}\rangle, \hskip 1cm \frac{d\Gamma^{frag}}{dz}=\sum_{a=q,\bar{q},g}\int_z^1\frac{dx}{x}C_a(x)D_{a\gamma}\left(\frac{z}{x},M\right),$$

where $C_a(x)$ represents the partonic kernels and $D_{a\gamma}(z/x, M)$ represents
the fragmentation functions, and $M$ is the mass of the vector quarkonium $J/\psi$. The partonic kernels can be expanded in powers of $v$. And $C_i(z,M)$ are short distance Wilson coefficients,
calculable in perturbation theory, while $\langle \Psi_V|\mathcal{O}_i|\Psi_V\rangle$ are the NRQCD matrix
elements with the $J/\psi$ quantum states $|\Psi_V\rangle$ ( $V=J/\psi$).

Many theoretical considerations have been done in this context \cite{tormo_inclusive_2007,field_phenomenological_2002}, and showed that, different approximations are needed in the different regions of the photon spectrum 
, and the spectrum has been described quite well in the case of bottomonium $\Upsilon$. While in the case of charmonium $J/\psi$ this NRQCD formalism is not able to describe the experimental data \cite{,tormo_inclusive_2007,cleo_collaboration_inclusive_2008}, which infers that the gluon mass still needs to be incorporated especially in the upper end-point region of the spectrum as suggested in \cite{cornwall_dynamical_1982,parisi_low_1980,field_phenomenological_2002} and advised also in \cite{tormo_inclusive_2007}.

The complete formula of the spectrum up to NLO \cite{i_tormo_semi-inclusive_2005} can be written as
$$\frac{d\Gamma}{dz}=\frac{d\Gamma_{LO}}{dz}+\frac{d\Gamma_{NLO}}{dz}+\frac{d\Gamma_{LO,\alpha_s}}{dz}$$
the first term consists of the gluon mass correction factors through the formula obtained by Liu and Wetzel \cite{liu_gluon-mass_1996}
\begin{equation*}\label{1}
\begin{split}
    \frac{1}{\Gamma_0}\frac{d\Gamma}{dz}=&\frac{1}{\pi^2-9}[\frac{x_+-x_-}{z^2}+\frac{\ln(x_+/x_-)}{z^2(-2+4\eta+z)^3}[8(2\eta-1)^2\\
&(2-4\eta+7\eta^2)+8(2\eta-1)(5-12\eta+10\eta^2+2\eta^3)z\\
& +2(2\eta-1)(-17+10\eta+6\eta^2)^2+2(-5+2\eta+2\eta^2)z^3]\\
& +\frac{(1/x_-1/x_+)}{z^2(-2+4\eta+z)^2}[4(2\eta-1)^2(1+3\eta^2)+4(2\eta-1)\\
& (3-4\eta+2\eta^2+2\eta^3)z+2(7-18\eta+10\eta^2+10\eta^3)z^2\\
& +4(2+\eta)(2\eta-1)z^3+(2+\eta)z^4]]
\end{split}
\end{equation*}
where $ z_{\pm}=1-2\eta-\frac{z}{2}\left[1\mp\sqrt{1-\frac{4\eta}{1-z}}\right]$ and $\eta=2m_g/M_{j/\psi}$, while the second and the third terms are detailed in refs \cite{i_tormo_semi-inclusive_2005,tormo_inclusive_2007}.


Our initial results of the gluon mass are approximate $m_g \approx 0.67\pm0.18$ GeV, which is simulated with a method similar to one in \cite{field_phenomenological_2002} by fitting the improved photon spectrum with the theoretical corrections given in \cite{tormo_inclusive_2007,i_tormo_semi-inclusive_2005} with the experimental data obtained by CLEO collaboration \cite{cleo_collaboration_inclusive_2008}. However, the result is not worth discussing here as it still needs more enhancements to reproduce the photon spectrum since the errors bar seems rather large.

\section{Search for hybrid mesons with exotic $J^{PC}$}
\label{sec:grate}
\noindent
The criteria to identify hybrid mesons such as X(1870) with quantum numbers $0^+0^{-+}$, $0^+1{^++}$, or $0^+2^{-+}$ as closest candidates are through the systematical study of its mass spectrum and strong decay properties, the results are not yet given here but we give an overview of calculations. However, the first-principle computations, directly from the QCD Lagrangian are extremely difficult, due to the failure of perturbation expansions for QCD at low energies. As a result, our calculations are based on the quark-gluon constituent model (QGCM) \cite{benhamida_hybrid_2020} as a QCD-based phenomenological potential model. This model requires two main external ingredients (the gluon mass $m_g$ and the strong coupling constant $\alpha_s$) that are extracted phenomenological (this work) or through continuum methods \cite{cornwall_dynamical_1982}.

\subsection{Mass spectrum}

In this case we shall use the following lowest-lying state $q\bar{q}$-cluster wave function for spin-space representation (for more details see \cite{benhamida_hybrid_2020}and references therein);
\begin{equation}
  \begin{split}
    \Psi_{JM}^{\textrm{PC}} (\overrightarrow{\rho},\overrightarrow{\lambda}) & = \left(\left(\left(\mathbf{e_{\mu_{g}}}\otimes\psi_{l_{g}}^{m_{g}}\right)_{j_{g}M_{g}}\otimes\psi_{l_{q\bar{q}}}^{m_{q\bar{q}}}\right)_{Lm}\otimes\chi_{\mu_{q\bar{q}}}^{S_{q\bar{q}}}\right)_{JM}^{\textrm{PC}}\\
      & = \sum_{\left(L\,l_{g}j_{g}l_{q\bar{q}}S_{q\bar{q}};\textrm{PC}\right)}\Psi_{JM;Ll_{g}j_{g}l_{q\bar{q}}S_{q\bar{q}}}(\overrightarrow{\rho},\overrightarrow{\lambda}),
  \end{split}
\end{equation}

Thus the hybrid meson mass can be calculated through.
\begin{equation}
M_{J^{PC}}\simeq En(M_i^0,\beta_{l_{q\bar{q}}l_g}^0)_{J^{PC}} + \Delta E(M_i^0,\beta_{l_{q\bar{q}}l_g}^0)_{J^{PC}}.
\end{equation}
 while
 $$ En(M_i^0,\beta_{l_{q\bar{q}}l_g}^0)_{J^{PC}}=\frac{\langle\Psi^{PC}_{JM}|H_R|\Psi^{PC}_{JM}\rangle}{\langle\Psi^{PC}_{JM}|\Psi^{PC}_{JM}\rangle},
\hskip 1cm\Delta E(M_i^0,\beta_{l_{q\bar{q}}l_g}^0)_{J^{PC}}=\frac{\langle\Psi^{PC}_{JM}| \Delta V_{SD}|\Psi^{PC}_{JM}\rangle}{\langle\Psi^{PC}_{JM}|\Psi^{PC}_{JM}\rangle}$$
where $H_R$ and $\Delta V_{SD}$ are the Hamiltonian of the hybrid meson state  and spin-dependent corrections term respectively  (we can not show all the details here since it will blow up the notation considerably, for more details the reader may take a look at ref \cite{benhamida_hybrid_2020}and references therein).
$$ H_R=\frac{\vec{p}^2_{\rho}}{2\mu_{\rho}}\frac{\vec{p}^2_{\lambda}}{2\mu_{\lambda}}+\frac{M_q}{2}+\frac{m_q^2}{2M_q} +\frac{M_{\bar{q}}}{2}+\frac{m_{\bar{q}}^2}{2M_{\bar{q}}}+\frac{M_g}{2}+\frac{m_g^2}{2M_g} + V_{eff}$$

\subsection{Strong decay}
To lowest order the decay of an hybrid state A to two light meson stats B and C is described by the matrix element of the Hamiltonian of quark pair creation
\begin{equation}
\left\langle BC|H|A\right\rangle =g\,f(A,B,C)\,\left(2\pi\right)^{3}\delta_{3}\left(\vec{p_{A}}-\vec{p_{B}}-\vec{p}_{C}\right).
\end{equation}
the Hamiltonian of the pair creation $g\rightarrow q\bar{q}$ is defined as
\begin{equation}
  H=\sum_{ss'\lambda}\int \frac{d\vec{p}d\vec{p'}d\vec{k}}{(2\pi)^3\sqrt{2\omega}}\delta^{(3)}(\vec{p}-\vec{p'}-\vec{k})\bar{u}_{\vec{p}s}\gamma_{\mu}\frac{\lambda^a}{2}u_{-\vec{p'}s'}b^{\dag}_{\vec{p}s}d^{\dag}_{-\vec{p'}s'}a_{\vec{k}\lambda}\varphi_a\varepsilon_{\vec{k}\lambda}^{\mu}.
\end{equation}
The partial width is given by:
\[
\Gamma_{A\rightarrow BC}=4\alpha_{s}\left|f(A,B,C)\right|^{2}\frac{P_{B}E_{B}E_{C}}{M_{A}},
\]

\section{Summary}
\label{sec:conc}

The advantage of the charmonium meson is that it is a multiscale system that can probe all regimes of QCD since its mass scale falls in between those low and high-energy scales. This makes it an excellent testing laboratory for the interplay between perturbative and non-perturbative QCD and an excellent scene for hunting the exotic mesons as well as the theoretical and phenomenological investigations of the gluon nature in the infrared, which may shed some light on the long-distance QCD dynamics since seemingly, gluons effective mass is a true QCD property.



\section*{Acknowledgments}
This work was supported by the PRFU research program (under No. B00L02EP310220190001).



\bibliographystyle{elsarticle-num}
\bibliography{citations} 

\end{document}